\documentclass[prb,twocolumn,a4paper,showpacs]{revtex4}
\usepackage{graphicx}

\begin{document}

\title{Dynamics of the Wang-Landau algorithm and complexity of rare events\\
 for the three-dimensional bimodal Ising spin glass}

\author{Simon Alder$^{1}$, Simon Trebst$^{1,2}$, 
        Alexander K. Hartmann$^{3}$, Matthias Troyer$^{1,2}$}
\affiliation{$^{(1)}$Theoretische Physik, Eidgen\"ossische Technische 
Hochschule Z\"urich, CH-8093 Z\"urich, Switzerland}
\affiliation{$^{(2)}$Computational Laboratory, Eidgen\"ossische
  Technische Hochschule Z\"urich, CH-8092 Z\"urich, Switzerland}
\affiliation{$^{3}$Institut f\"ur Theoretische Physik, Universit\"at
  G\"ottingen, 37077 G\"ottingen, Germany}

\date{\today}

\begin{abstract}
We investigate the performance of flat-histogram methods based on a multicanonical
ensemble and the Wang-Landau algorithm for the three-dimensional $\pm J$ spin 
glass by measuring round-trip times in the energy range between the zero-temperature 
ground state and the state of highest energy. 
Strong sample-to-sample variations are found for fixed system size and the
distribution of round-trip times follows a fat-tailed Fr\'echet
extremal value distribution. Rare events in the fat tails of these
distributions corresponding to extremely slowly equilibrating spin
glass realizations dominate the calculations of statistical
averages. 
While the {\em typical} round-trip time scales exponential as expected for this NP-hard
problem,  we find that the {\em average} round-trip time is no longer well-defined
for systems with $N \geq 8^3$ spins.  
We relate the round-trip times for multicanonical sampling to 
intrinsic properties of the energy landscape and compare with the 
numerical effort needed by the genetic Cluster-Exact Approximation
to calculate the exact ground state energies.
For systems with $N \geq 8^3$ spins the simulation of these rare events 
becomes increasingly hard. For $N \geq 14^3$ there are samples where the Wang-Landau 
algorithm fails to find the true ground state within reasonable simulation times.  
We expect similar behavior for other algorithms based on multicanonical sampling.

\end{abstract}


\maketitle


\section{Introduction}

The three-dimensional $\pm J$ Ising spin glass has been extensively
studied\cite{SpinGlassReview} as a prototype system which exhibits a
finite temperature second order phase transition to a slowly
equilibrating glassy phase.\cite{Ballesteros:00,Palassine:99} The
simulation of such a system with conventional Monte Carlo methods is
slowed down by long relaxation times in the spin-glass phase. This
problem has been addressed by a number of algorithmic developments
such as the multicanonical method,\cite{Multicanonical} simulated and
parallel tempering,\cite{Tempering} broad
histograms\cite{BroadHistograms} and transition matrix Monte
Carlo.\cite{TMMC} In order to speed up equilibration most of these
methods aim at broadening the energy range sampled within the Monte
Carlo (MC) simulations from the sharply peaked distribution of
canonical sampling at a fixed temperature.

Recently, Wang and Landau introduced a new algorithm which systematically calculates an
estimate of the density of states and iteratively converges to
sampling a flat histogram in energy.\cite{WangLandau} The Wang-Landau (WL) algorithm simulates a biased random walk in configuration space. The bias depends only on the total energy of a configuration and is defined by a statistical ensemble with weights $w(E)$. For this ensemble the transition probabilities are given by the Metropolis scheme
\begin{equation}
  p(E_1 \rightarrow E_2) = \min \left( \frac{w(E_2)}{w(E_1)}, 1\right) \;.
\end{equation}  
The equilibrium distribution of the energy in this ensemble is $n_w(E) \propto w(E)g(E)$ where $g(E)$ is the density of states. By setting the weights $w(E) \propto 1/g(E)$ the WL algorithm aims at sampling a flat histogram in energy.
The crucial feature of the WL algorithm however is that the simulated ensemble is {\em dynamically} modified during the course of the simulation: after every spin update the current estimate of the density of states is multiplied by a modification factor $f$ and the ensemble weights are analogously updated. The modification factor is iteratively reduced to 1 whenever the sampled energy histogram is close to the expected equilibrium distribution, that is when the histogram is ``flat" within a given range.
The dynamic modification of the ensemble allows to push the random walker towards the low entropy states in the initial stages of the algorithm while ensuring that it converges to a flat-histogram/multicanonical ensemble in the final stages of the computation.

In this paper, we determine the performance of the Wang-Landau
algorithm for the three-dimensional $\pm J$ Ising spin glass for both
stages of the algorithm.   
First, we study the {\em dynamic behavior} of the
algorithm in the initial stages by considering its ability to find the
ground-state energy of a number of three-dimensional spin glass
samples which is a well known NP-hard problem.\cite{Barahona:82} 
We compare the obtained ground state energies to exact results calculated 
with the genetic Cluster-Exact approximation (CEA) \cite{Hartmann,Hartmann:Book}.
While the WL algorithm reproduces the exact ground-state energy for small systems,
we find that for moderately large systems $(N \geq 14^3)$ the WL
algorithm does not find the exact ground-state energy for few 
spin-glass samples. Even when restricting the simulated energy bin around
the known ground-state energy the algorithm does not find the lowest
energy state within a reasonable number of sweeps $(N_{\mbox{sweeps}}
\approx 10^7)$.  
The genetic CEA gives a superior performance to find ground state energies, 
but does not give the full thermodynamic information.
Second, we investigate the {\em asymptotic behavior} of the WL algorithm 
by measuring round-trip times in energy for the converged ensemble. 
The round-trip time gives a direct estimate of the equilibration time for the 
multicanonical ensemble. The asymptotic scaling of the Wang-Landau algorithm 
therefore also reflects the performance of other flat-histogram methods based on 
multicanonical sampling, such as the multicanonical method,\cite{Multicanonical}
broad histograms\cite{BroadHistograms} or transition matrix Monte
Carlo.\cite{TMMC} We find large sample-to-sample variations of the
round-trip times which can be described by fat-tailed Fr\'echet
extremal-value distributions. We discuss important implications for
statistical sample averaging which are caused by the rare events in
the fat tails of these distributions. The intrinsic character of the
observed extremal-value distributions is demonstrated. Finally it is
shown that these distributions scale exponentially with the linear
system size.  
In comparison, we find that the computational effort of the genetic CEA is also
correlated to the density of states, but is less sensitive to the low-energy landscape 
than the WL algorithm.
Finally, we discuss these limitations of multicanonical sampling by measuring
the local diffusivity of the random walker in energy. We find a
pronounced minimum of the diffusivity near the ground-state energy
which is symptomatic for the entropic barrier which slows down the
equilibration of the random walker.


\section{Dynamic Performance}

To study the dynamic performance of the Wang-Landau algorithm we have
tested its ability to find the ground-state energy of a number of spin
glass samples where the exact ground state is known.  Finding the
ground-state energy of the $\pm J$ spin-glass is an extensively
studied problem, both by physicists and computer scientists. For the
two-dimensional $\pm J$~spin glass the \emph{full} density of states
can be calculated with polynomial
effort.\cite{Bieche:80,SaulKardar:94} However, for the
three-dimensional $\pm J$~spin glass as well as the two-dimensional
$\pm J$~spin glass with an external field, even the problem of finding
the ground-state energy has been shown to be
NP-hard.\cite{Barahona:82} The direct calculation of the ground-state
energy of spin glass samples using sophisticated exact branch-and-cut
algorithms \cite{simone95,simone96} is therefore limited to rather small systems.  
In this study, we applied a combination of  a genetic algorithm with 
Cluster-Exact Approximation (CEA) to determine ground state energies.
\cite{Hartmann,Hartmann:Book} 
The algorithm, although based on heuristics, is able to find true ground states of
samples up to size $N=16^3=4096$ in reasonable, although exponentially
growing time \cite{Hartmann:Book}

We use the results obtained from this genetic + CEA approach to check the 
accuracy of the ground-state energies found with the WL algorithm for samples
up to size $N=14^3$.  For system size, $N \leq 6^3$, the density of
states of each sample was calculated without energy binning by $50$
independent runs.  For all runs we find that the WL algorithm gives
the exact ground-state energy. For $N=8^3$, for some few samples 
(17 out of 1000) the exact ground-state energy was found only by running
extensive runs after the comparison with the results from the heuristic approach
revealed that the true ground state energy has not been found.

For larger systems with $N=10^3$ and more spins we have restricted the WL simulations to the smallest allowed energy range around the ground-state energy calculated by the genetic CEA. 
In order to keep ergodicity the energy bin has to be larger than 
\begin{equation}
  \Delta E / J > 4L^{d-1} \;,
  \label{eq:bin_size}
\end{equation}
to assure that two domain walls can be inserted which thus enables moves which subsequently flip all spins and ergodicity within the energy bin is given.

Each sample was simulated by six independent runs. 
For all samples with $N=10^3$ spins the WL algorithm found true 
ground-state energies. However, for three samples not all runs gave the true ground-state energy, but sampled energies down to the first excited state only (within $(2.8 \pm 0.2) \times 10^7$ MC sweeps). 
For the samples with $N=12^3$ spins we find similar results. For all samples there is at least one run which finds the true ground-state energy within some $(2.9 \pm 0.1) \times 10^7$ MC sweeps.
However, for 9 out of 10 samples the WL algorithm does not converge to sampling a flat histogram in the full energy range.
For the samples with $N=14^3$ spins the WL algorithms finds the true ground-state energy for 4
samples only. For 6 out of 10 samples the algorithm did not sample the
exact ground-state energy once within $(3.0 \pm 0.1)\times 10^7$ MC
sweeps for all independent runs. Furthermore, there is one sample
where even the first excited state is not found within the given
number of sweeps. For all samples the simulated ensemble does not
converge towards the multicanonical ensemble sampling a flat energy
histogram and the WL algorithm gets stuck.


\section{Asymptotic Performance}
\label{AsymptoticPerformance}

\begin{figure}
  \includegraphics[scale=0.35]{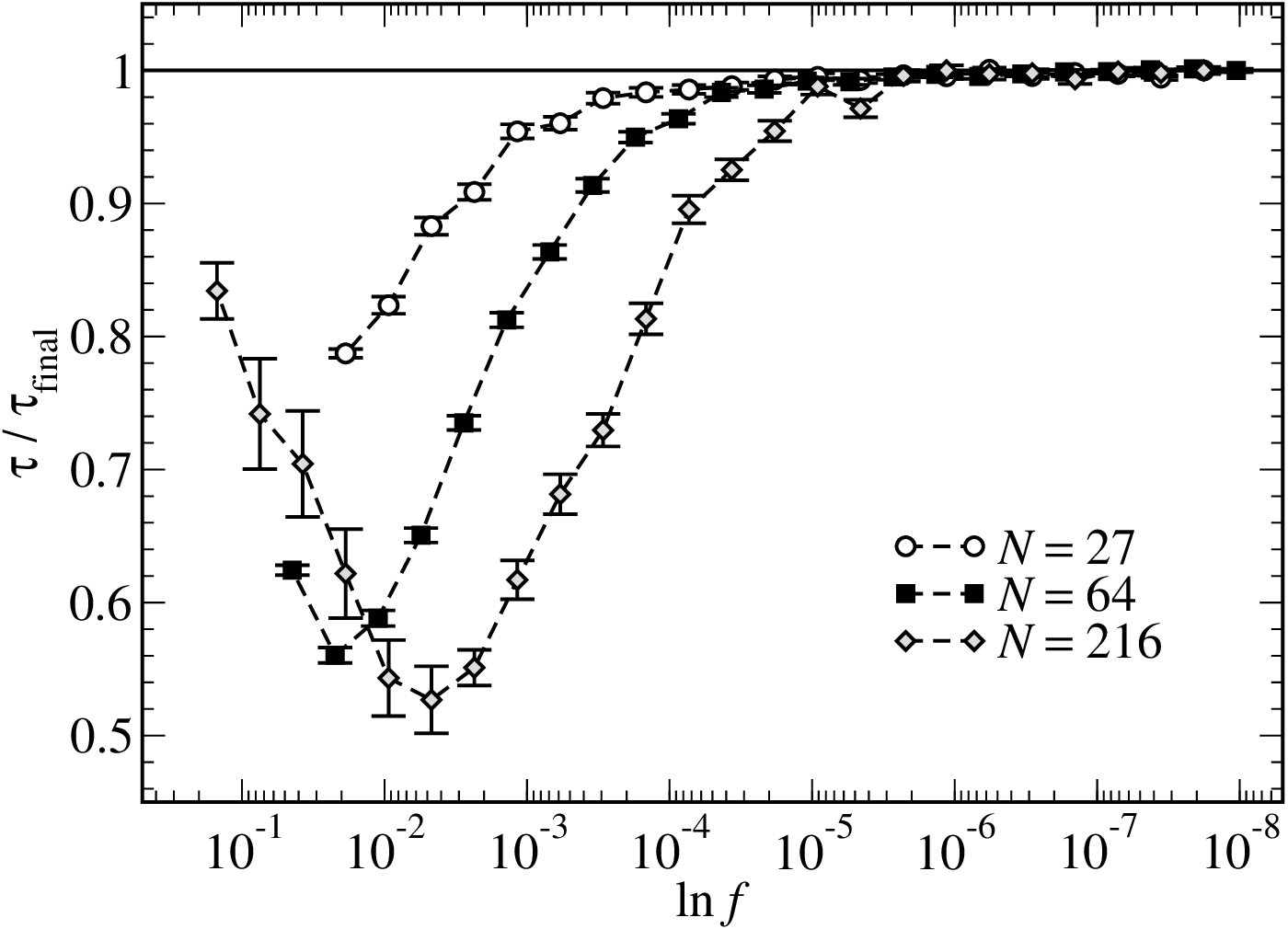}
  \caption{ 
    Convergence of the round-trip time $\tau$ versus the modification factor $f$ in the Wang-Landau
    algorithm. Shown are the results for three randomly generated three-dimensional $\pm J$~spin 
    glass samples with $N=3^3=27, 4^3, 6^3=216$ spins. The measurements are averaged over $500$ 
    independent runs for $N=27$, $64$ and over $30$ runs for
    $N=216$, respectively. }
  \label{fig:round_trip_time_convergence}
\end{figure}

We now turn to the asymptotic behavior of the WL algorithm and
flat-histogram methods in general which we determine by measuring the
round-trip times in energy of the simulated random walker for the
converged flat-histogram ensemble. Here the round-trip time
corresponds to the number of single-spin flips needed to get from a
configuration with the ground-state energy to a configuration with
highest energy (the anti ground state). The round-trip time thus gives an estimate of the
equilibration time for the flat-histogram ensemble.  For the Ising
model the number of energy levels scales linearly with system size
$N$. While the round-trip time of an unbiased random walker scales
like $\tau \sim N^2$, it was recently shown that for various
two-dimensional Ising models the growth with the number of spins is
significantly stronger for the biased flat-histogram random
walker.\cite{Dayal:04}  For the ferromagnetic and fully frustrated
Ising models polynomial scaling, $\tau \sim N^{2.4}$ and $\tau \sim
N^{2.9}$, was found, and exponential growth for the two-dimensional
$\pm J$ spin glass.\cite{Dayal:04}

For a given sample we find that the round-trip time measured during the iterations of the WL algorithm converges as the simulated ensemble approaches the flat-histogram ensemble. The convergence of round-trip times is illustrated for three randomly generated spin glass samples in Fig.~\ref{fig:round_trip_time_convergence}. Since correct convergence to the round-trip times of the {\em exact} flat-histogram ensemble was shown for the two-dimensional $\pm J$ Ising spin glass\cite{Dayal:04} we assume correct convergence for the 3D case as well and thereby justify that our results for the asymptotic round-trip times hold for {\em any} flat-histogram method.

\subsection{Sample-to-sample variation}

\begin{figure}
  \includegraphics[scale=0.35]{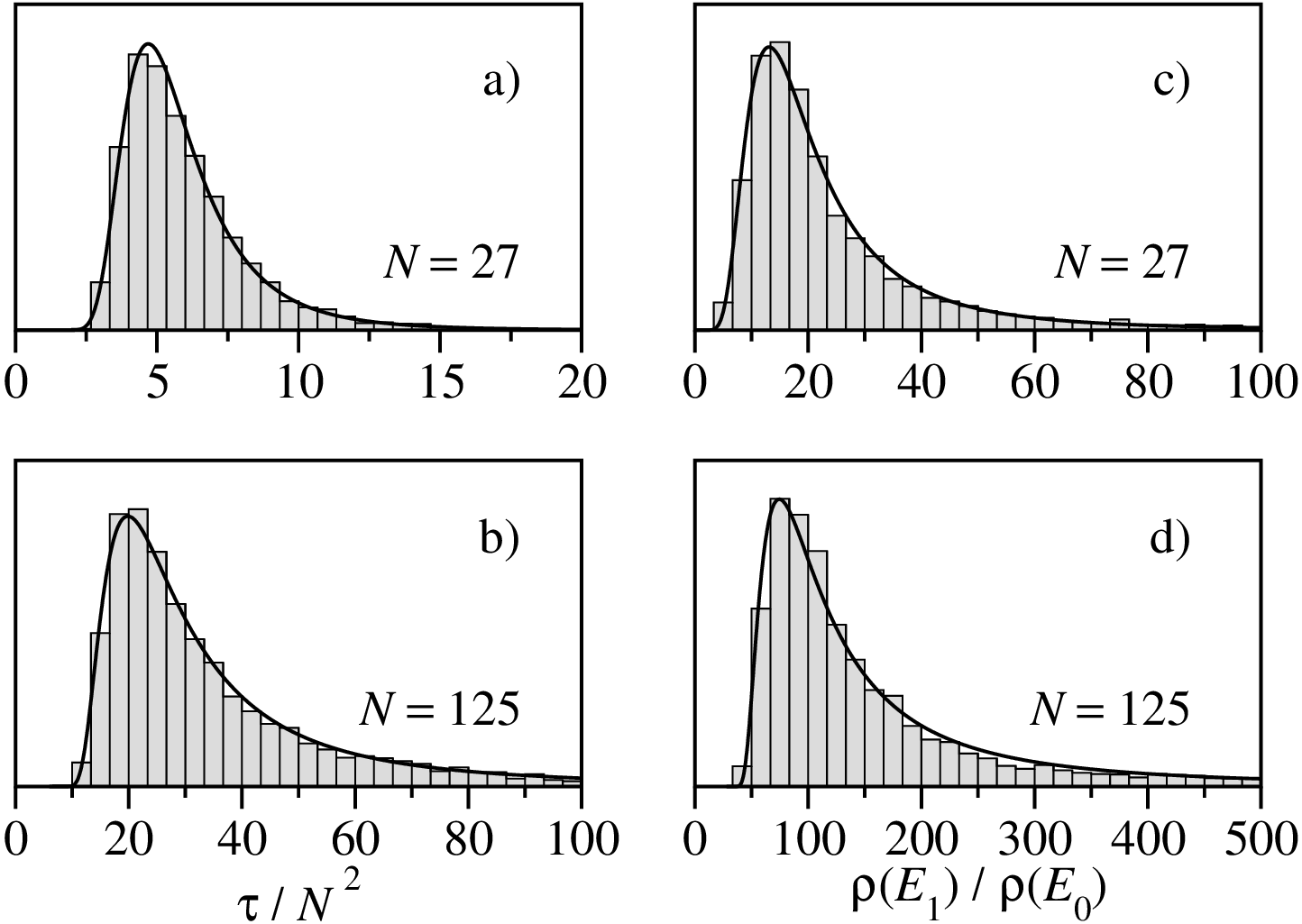}
  \caption{
    Left panels:  Distribution of round-trip times $\tau$ for 5000 randomly generated spin glass
    samples of size $N=3^3=27$ and $N=5^3=125$ respectively.
     Right panels: Distribution of the ratio of the number of first excited states to the number of 
     ground states $g(E_1)/g(E_0)$ for the same system sizes.
    In all panels the solid lines indicate fits to fat-tailed Fr\'echet extremal-value distributions.
  }
  \label{fig:histogram_tau_rho}
\end{figure}

\begin{figure}
  \includegraphics[scale=0.35]{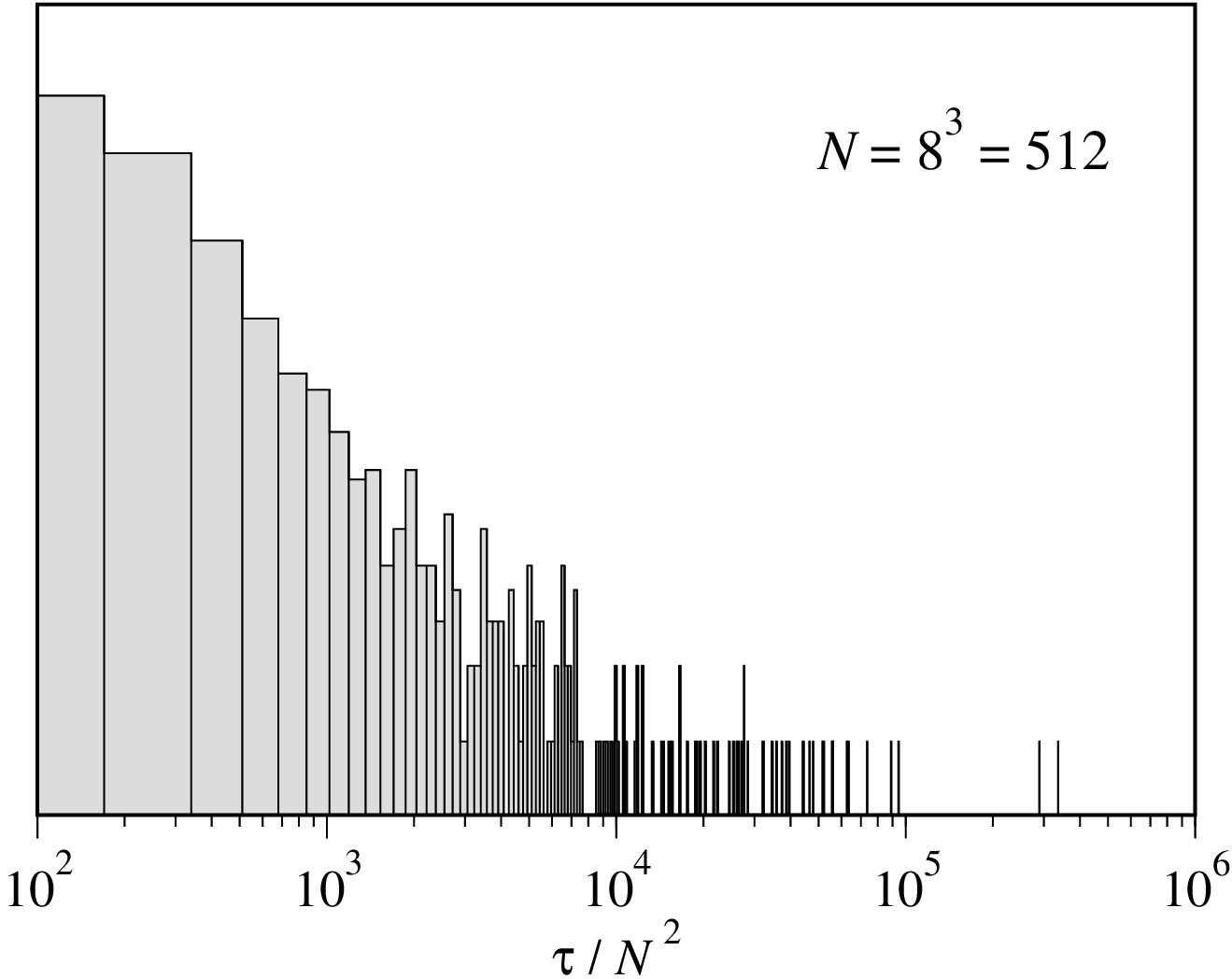}
  \caption{    
    Log-log plot of the distribution of round-trip times $\tau$ for the three-dimensional $\pm J$~spin 
    glass with $N=8^3=512$ spins. The round-trip times $\tau$ were measured for the converged 
    flat-histogram ensemble in the final step of the Wang-Landau algorithm. 
    Data from $1000$ randomly generated spin glass samples are shown. 
  }
  \label{fig:histogram_8}
\end{figure}

To study the sample dependence of the round-trip times we have
analyzed $5000$ randomly generated spin glass samples for $N=3^3$,
$4^3$, $5^3$, $6^3$ and $1000$ samples for $N=8^3=512$,
respectively. To assure convergence of the measured round-trip times
we restrict the measurement to the final step in the Wang-Landau
algorithm. We find strong sample-to-sample variations over several
orders of magnitude for fixed system size $N$ which is shown in the
left panels of Fig.~\ref{fig:histogram_tau_rho} ($N=3^3$ and
$N=5^3$). 
For the spin glass of size $N=8^3=512$ the full distribution of round-trip times is shown in 
Fig.~\ref{fig:histogram_8}. The distribution covers some 4 orders of magnitude and contains spin glass samples which were simulated between some minutes and about a month on a $500$~MHz Pentium~III CPU.

To quantitatively analyze the extremal events in the tails of the distributions, we use extremal-value theory.\cite{Extremal1} 
The central limit theorem for extremal-value states that the extrema of random subsets of any distribution follow a generalized extremal-value distribution.\cite{Extremal2} This distribution takes one of three characteristic forms: Fr\'echet (algebraic decay, fat-tailed), Weibull (exponential decay) or Gumbel (faster than exponential decay, thin-tailed).
Here we find that  \emph{all measured round-trip times} seem to follow a fat-tailed Fr\'echet extremal-value distribution, similar to the two-dimensional $\pm J$~spin glass.\cite{Dayal:04}  This implies that the central limit theorem applies even to the smallest possible subset -- a  {\em single} round-trip time. As a consequence, every single three-dimensional $\pm J$~spin glass sample constitutes an extremal event.
The integrated form of the Fr\'echet distribution is given by
\begin{equation}
  H_{\xi ,\mu ,\beta}(\tau) =
  \exp \left( - \left( 1 + \xi \frac{\tau - \mu}{\beta}
  \right)^{-1/\xi} \right) \, .
  \label{eq:frechet}
\end{equation}
The parameter $\mu$ indicates the location of the distribution, that is the most probable round-trip time, and the parameter $\beta$ defines the scale of the distribution, e.g. the height of the peak. The shape parameter $\xi$, which is positive for fat-tailed distributions, describes the decay of the tail of the distribution. We have determined the three parameters $\mu$, $\beta$ and $\xi$ of the fitted Fr\'echet distributions with a maximum likelihood estimator. The resulting fits are shown as solid lines in Fig.~\ref{fig:histogram_tau_rho}.

\begin{figure}
  \includegraphics[scale=0.35]{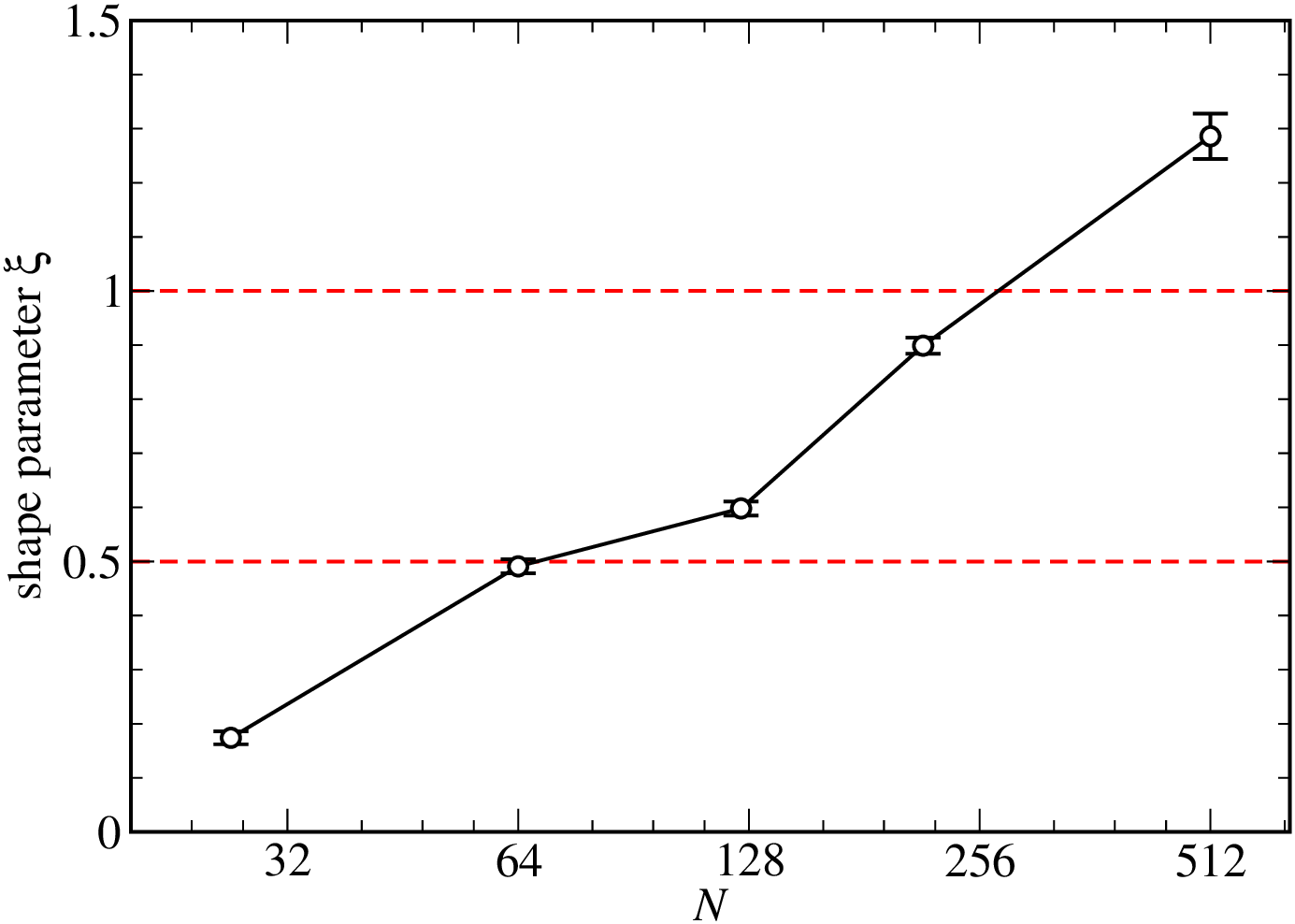}
  \caption{    
    Scaling of the shape parameter $\xi$ of the fitted Fr\'echet distributions versus system size.
    The $m$-th moment of a fat-tailed Fr\'echet distribution is well-defined only if $\xi < 1/m$. 
    The dashed lines indicate where variance and mean of the respective distributions become 
    ill-defined.
    }
  \label{fig:scaling_xi}
\end{figure}

We now turn to the scaling of the shape parameter with system size $N$ which is shown in  Fig.~\ref{fig:scaling_xi}. With increasing system size the shape parameter monotonically increases. 
This strongly affects the fat tails of the distributions which in the limit $\tau \to \infty$ exhibit a power-law decay of the form
\begin{equation}
  \frac{\text{d}}{\text{d}\tau}H_{\xi , \mu , \beta}(\tau)
  \stackrel{\tau\to\infty}{-\!\!\!\!-\!\!\!\!-\!\!\!\!-\!\!\!\!\longrightarrow}
  \tau^{-\left( 1+1/\xi
  \right)} \, .
  \label{eq:H_xi}
\end{equation}
From this asymptotic behavior we can see that the $m$-th moment of a fat-tailed Fr\'echet distribution is well-defined only if $\xi < 1/m$. 
The scaling in Fig.~§\ref{fig:scaling_xi} suggests that
for $N > 4^3=64$ ($N \geq 8^3=512$) the shape parameter becomes larger than $0.5$ ($1$) and thus the variance (mean) of the distribution is no longer well-defined. 

\begin{figure}
  \includegraphics[scale=0.35]{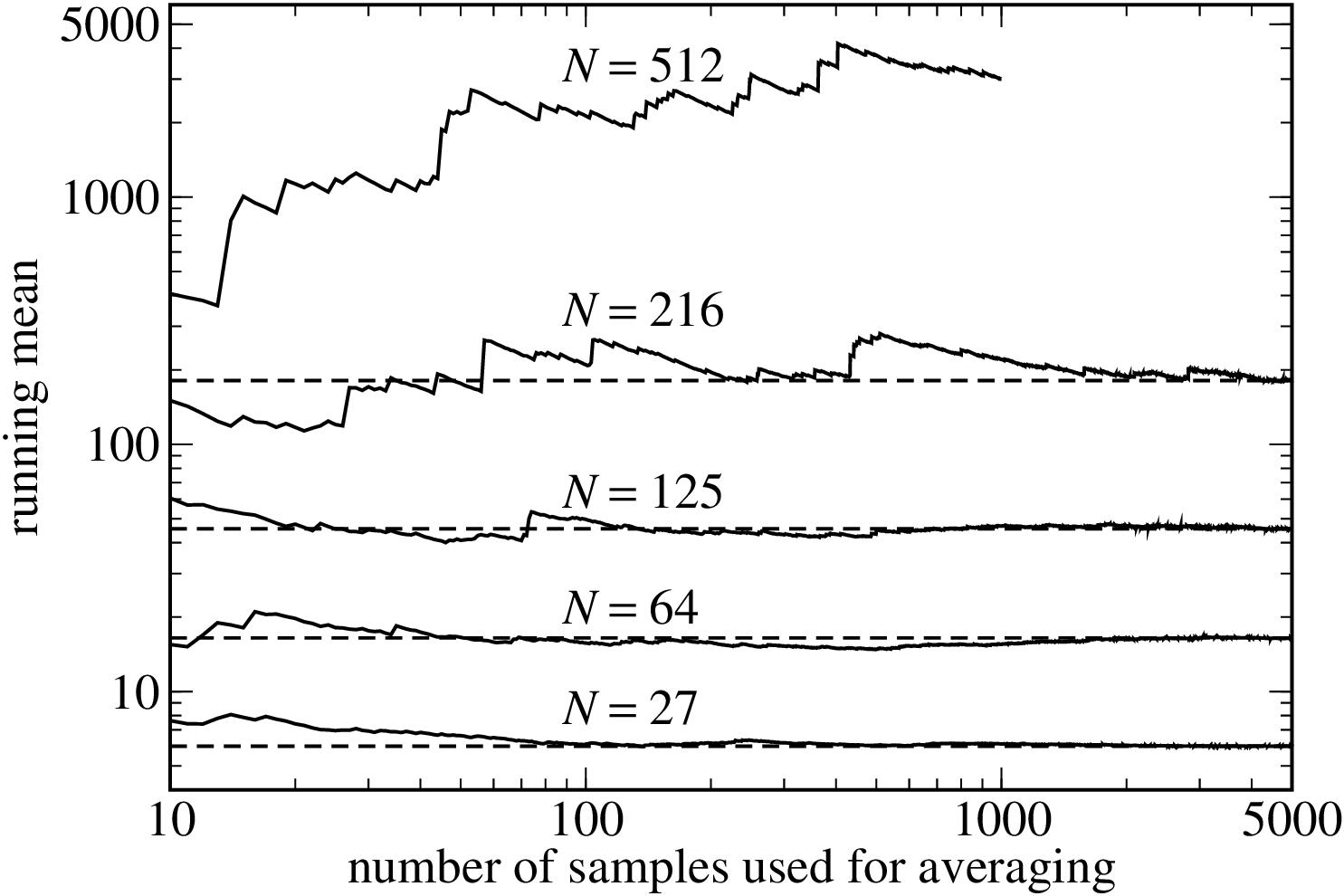}
  \caption{
    Log-log plot of the running mean of round-trip times defined in Eq.~(\ref{eq:mean}) versus the 
    number of samples used for averaging. Results for various system sizes are shown. 
    For $N \leq 6^3=216$ the mean is well-defined and the running mean seems to converge  
    (dashed line). For $N=8^3=512$ the mean becomes ill-defined and the running mean diverges
    \cite{footnote}.
  }
  \label{fig:running_mean}
\end{figure}

\begin{figure}
  \includegraphics[scale=0.35]{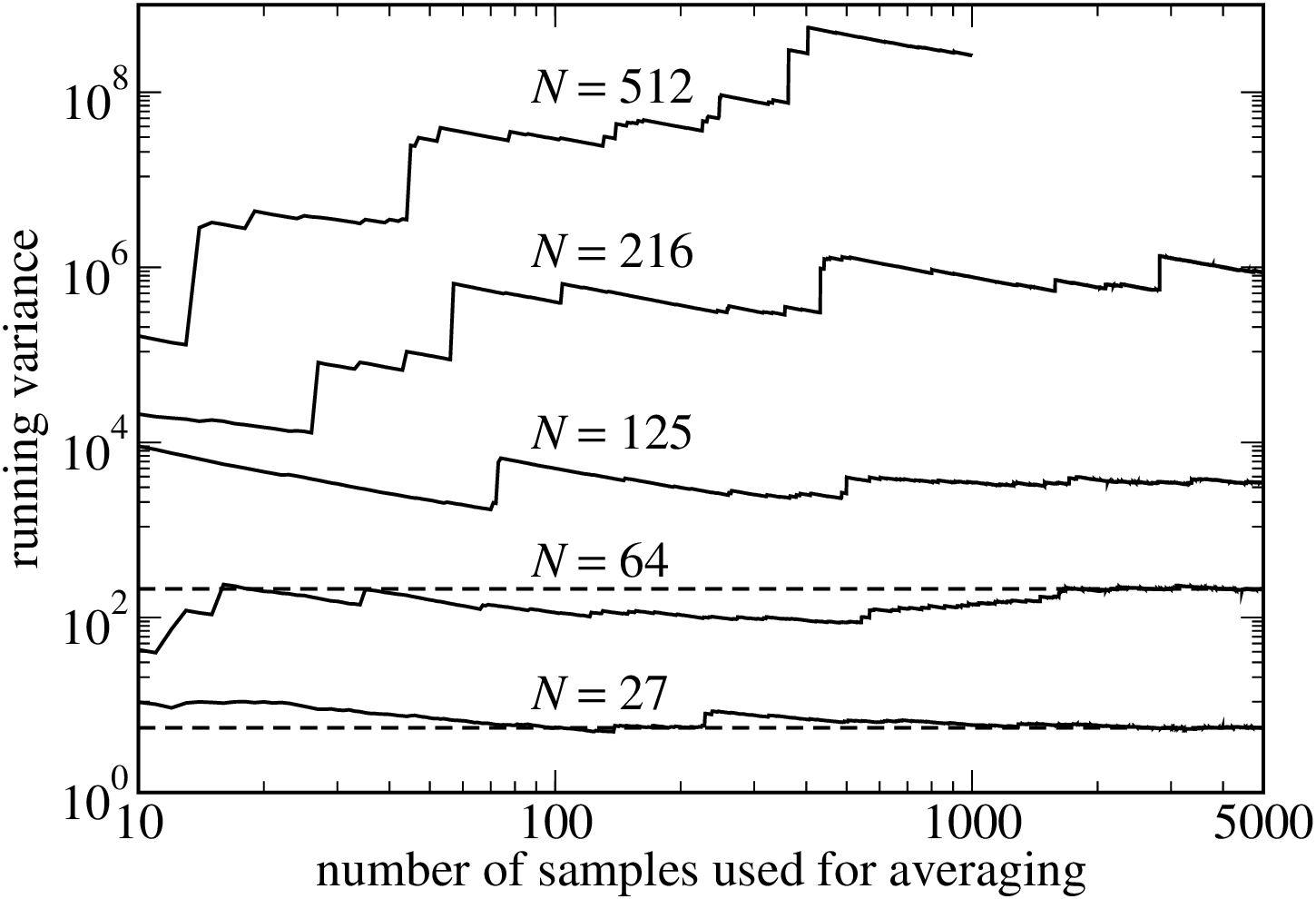}
  \caption{
    Log-log plot of the running variance of the distribution of round-trip times defined in 
    Eq.~(\ref{eq:variance}). Results for various system sizes are shown. 
    For $N \leq 4^3=64$ the variance is well-defined and the running variance seems to converge 
    (dashed line). For $N \geq 5^3=125$ the variance becomes ill-defined and the running variance 
    diverges \cite{footnote}. 
  }
  \label{fig:running_variance}
\end{figure}

To illustrate this unusual behavior we can calculate running moments of the distribution, e.g. by only considering subsets of the first $n$ round-trip times of all measured round trip times $\{ \tau \}$ when calculating the moments of the distribution.
The running mean of round-trip times is then defined by
\begin{equation}
  \text{Mean}_{\{\tau\}}(n) = \frac{1}{n}\sum_{i=1}^n \tau_i \;,
  \label{eq:mean}
\end{equation}
and the running variance by
\begin{equation}
  \text{Var}_{\{\tau\}}(n) = \frac{1}{n-1}\sum_{i=1}^n \left(\tau_i - \text{Mean}_{\{\tau\}}(n) \right)^2\;.
  \label{eq:variance}
\end{equation}
The running mean and variance are calculated for a fixed random order of the set of round-trip times $\{ \tau \} \equiv \{ \tau_1 , \tau_2, \ldots , \tau_{5000} \}$. 
Figs. ~\ref{fig:running_mean} and \ref{fig:running_variance} show the running mean and variance for various system sizes. For small system sizes, $N \leq 64$, both mean and variance are well defined and the running mean and its error converges. For larger systems the variance becomes ill-defined as the shape parameter $\xi$ turns larger than $0.5$.  Irregular ``jumps"  indicating rare events occur in the calculation of the running mean. For systems with $N \leq 216$ spins the running mean still converges, but the error of the running mean does not reduce even for large sample sets.
For systems with more than $N=8^3=512$ spins the shape parameter becomes larger than $1$ and the mean thus ill-defined. This divergence of the mean round-trip time also becomes apparent in the calculation of the running mean where the irregular jumps occur so frequently that the mean round-trip time no longer converges \cite{footnote}.

This behavior becomes even more evident for the variance, the second moment of the distribution, which according to the scaling of the shape parameter illustrated in Fig.~\ref{fig:scaling_xi} is no longer well-defined for systems larger than $N > 4^3=64$. 
As illustrated in Fig.~\ref{fig:running_variance} the running variance diverges for larger systems
\cite{footnote}. Again, frequent ``jumps" in the running variance indicate the occurrence of rare events in the fat tails of the distribution which dominate the calculation of the running variance.

\subsection{Intrinsic correlations for the WL algorithm}

\begin{figure}
  \includegraphics[scale=0.325]{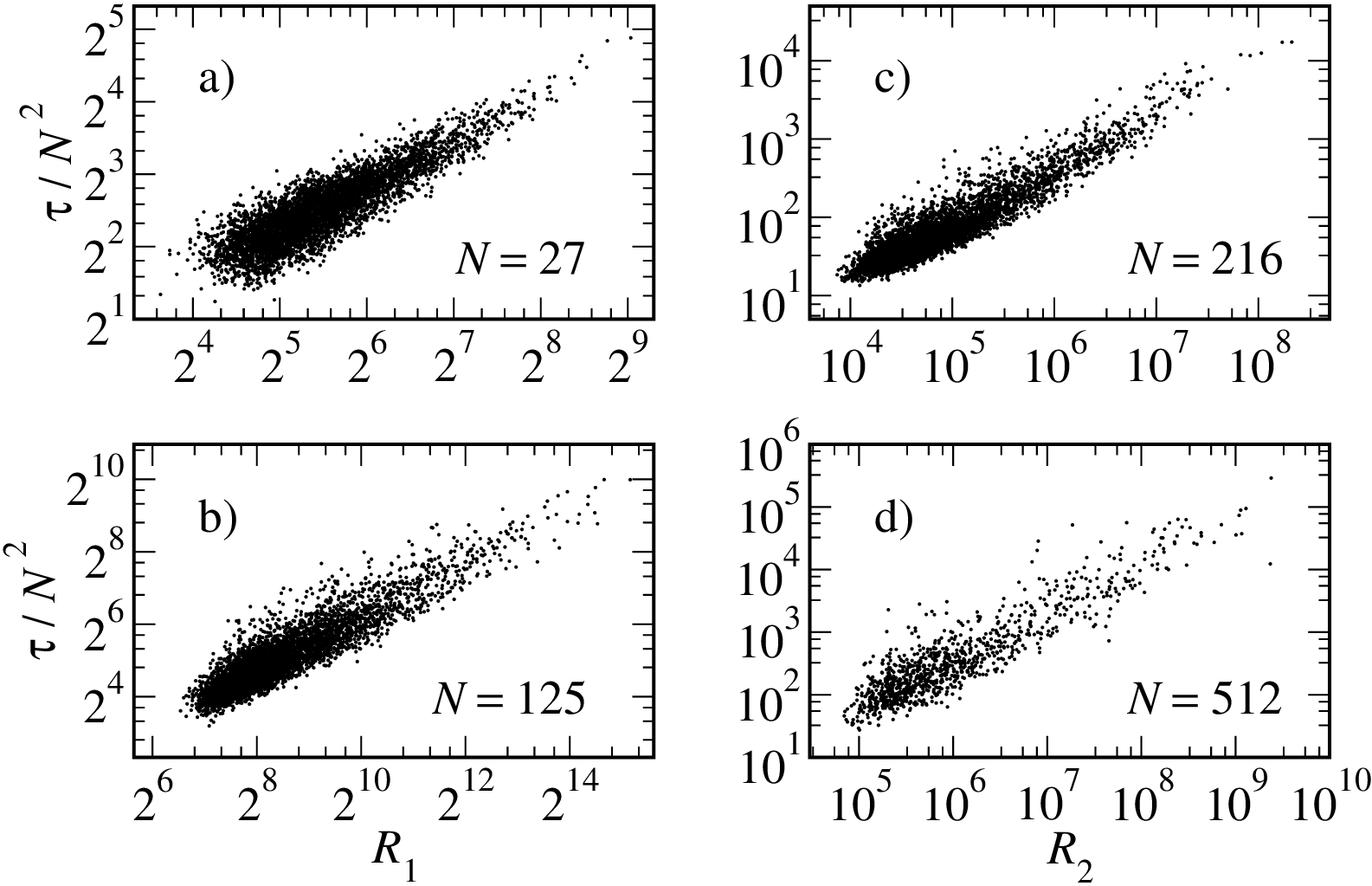}
  \caption{
    Correlation of the round-trip time $\tau$ and ratios in the density of states 
    for various system   sizes. 
    In the left panels the ratio $R_1 = g(E_1)/g(E_0)$ is shown. 
    In the rights panels transitions to higher excited states are also included, 
    $R_2 = g(E_1)/g(E_0) + g(E_2)/g(E_1) + g(E_2)/g(E_0)$.
    Shown are data from 5000 randomly generated spin glass samples for
    $N=27$, $125$, $216$ and 1000 for $N=8^3=512$, respectively.}
  \label{fig:correlation}
\end{figure}

    In order to test whether the occurrence of the Fr\'echet
    extremal-value distributions reflect an intrinsic property of the
    energy landscape 
    three-dimensional spin glass we have analyzed the calculated
    density of states near the ground-state energy. The energy
    landscape near the ground state dominates to a large extent the
    measured round-trip times which is further discussed in the
    context of diffusivity measurements in section \ref{Diffusivity}.
    Here we study ratios in the density of states, such as the number
    of first excited states to the number of ground states,
    $g(E_1)/g(E_0)$. From the ground state a single-spin-flip update
    can connect at most $N g(E_0)$ states of energy $E_1$ and thus the
    ratio $g(E_1)/g(E_0)$ gives a qualitative measure of the number of
    local minima, i.e. which are not reachable from a ground state via
    a single spin flip.  In the right panels of
    Fig.~\ref{fig:histogram_tau_rho} the distribution of this ratio
    $g(E_1)/g(E_0)$ is shown for 5000 spin glass samples of size
    $N=3^3=27$ and $N=5^3=125$ respectively. Again we find that these
    distributions follow fat-tailed Fr\'echet extremal-value
    distributions. For these system sizes a strong correlation to the
    distribution of round-trip times is found which spans over several
    orders of magnitude as demonstrated in the left panels of
    Fig.~\ref{fig:correlation}.  For larger systems these correlations
    become less pronounced. However, if we consider additional
    transitions, such as the transition from the second to first
    excited state, $E_2 \to E_1$ and $E_2 \to E_0$, and calculate the
    sum of the respective ratios in the density of states we can
    recover a correlation over several orders of magnitude as it is
    shown for systems with $N=6^3=216$ and $N=8^3=512$ spins in the
    right panels of Fig.~\ref{fig:correlation}.

\subsection{Intrinsic correlations for the heuristic approach}

\begin{figure}
  \includegraphics[scale=0.325]{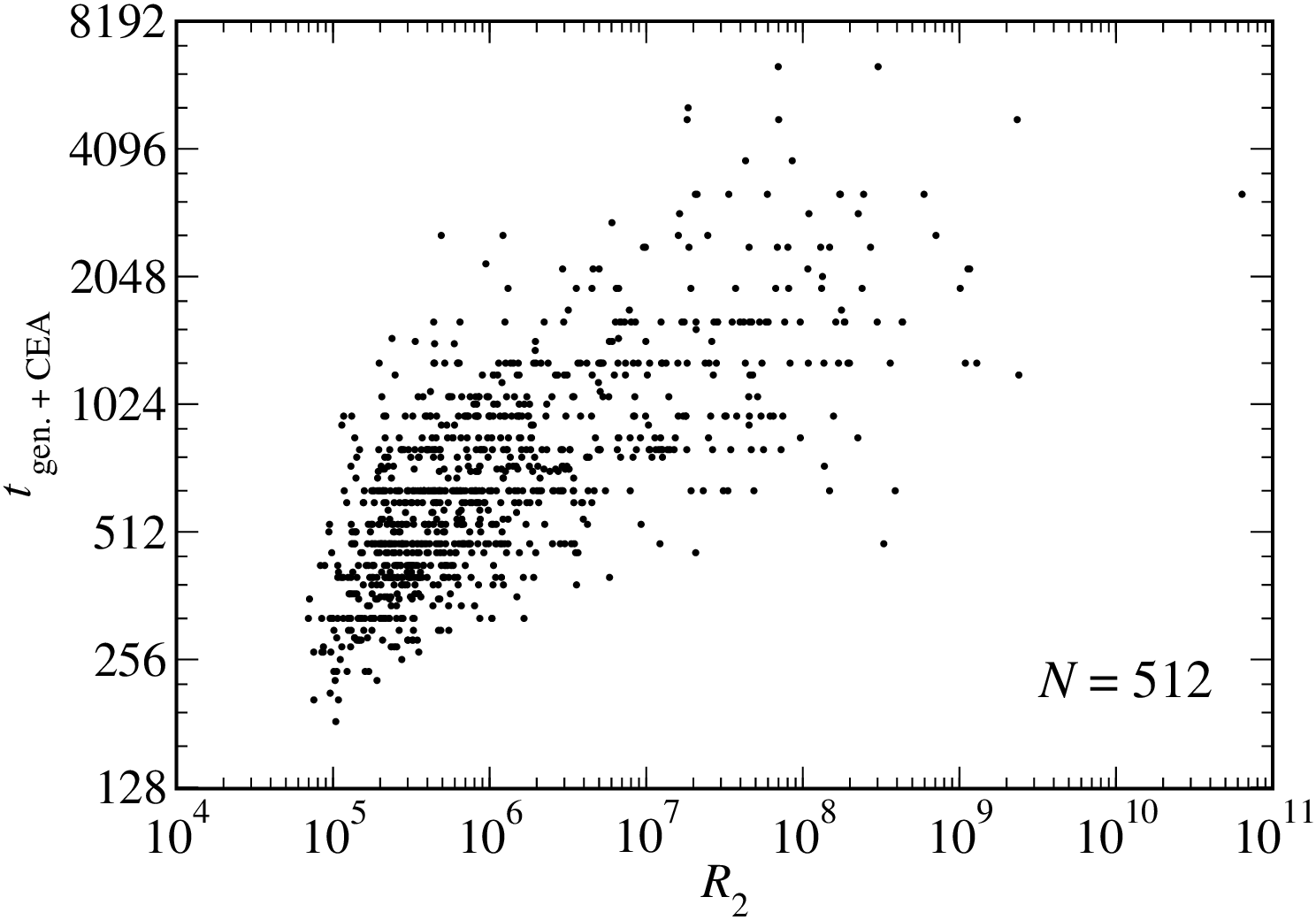}
  \caption{
    Correlation of the computational effort, $t_{\rm gen. + CEA}$, of the genetic CEA (see text) 
    and the ratio in the density of states for $N=8^3$ (1000 samples). 
    Here the ratio $R_2 = g(E_1)/g(E_0) + g(E_2)/g(E_1) + g(E_2)/g(E_0)$ is shown.
   The correlation to $R_1 = g(E_1)/g(E_0)$ looks similar.}
  \label{fig:correlationCEA}
\end{figure}

The strong correlation between intrinsic features of the energy landscape
and the measured round-trip times for flat-histogram sampling naturally leads
to the question whether the computational effort of other algorithms also 
comply with these intrinsic features. 
To this end, we compare the computational effort of the alternative heuristic
approach with the density of states and the round-trip times measured in the
WL algorithm for 1000 samples with $N=8^3=512$ spins.

We define the computational effort of the genetic CEA as the running time
of the algorithm which strongly depends on the parametrization of the underlying
genetic algorithm, namely
\begin{itemize}
\item $M_{\rm i}$ = initial size of population, i.e. how many configurations
are optimized in parallel
\item $n_{\rm o}$ = average number of offspring 
per configuration, i.e. how many iterations of the algorithm are run
\item $n_{\min}$ = number of CEA minimization steps per configuration
  and per iteration
\end{itemize}
The running time then depends linearly on the product  $M_{\rm i}n_{\rm o}n_{\min}$.
If the algorithm is run independently several times, one finds that not all runs result 
in ground states and we denote the fraction of runs which find the true ground state as 
$f_{\rm GS}$. In general one can observe that this fraction increases with the running time.

We now define the computational effort for a given sample and parameter set 
$\{M_{\rm i},n_{\rm o},n_{\min}\}$ as
$\tilde{t}(M_{\rm i},n_{\rm o},n_{\min}) = M_{\rm i}n_{\rm o}n_{\min}/f_{\rm GS}$ 
(i.e. $t=\infty$, if no ground state is found). 
For ``simple'' samples, many ground states are found for most parameter combinations, 
while ``hard'' samples need large sizes of populations and/or many iterations
and/or many minimization steps, thereby increasing the computational effort. 
For a given spin glass sample we perform multiple simulations with different combinations of parameters, and define the overall computational effort of a sample as the minimum over all
parameter combinations considered (using a fixed value $n_{\rm R}=20$):
\begin{equation}
t_{\rm gen. + CEA}=\min_{(M_{\rm i},n_{\rm o},n_{\min})} 
\left( M_{\rm i}n_{\rm o}n_{\min}/f_{\rm GS} \right)\,.
\end{equation}

\begin{figure}
  \includegraphics[scale=0.325]{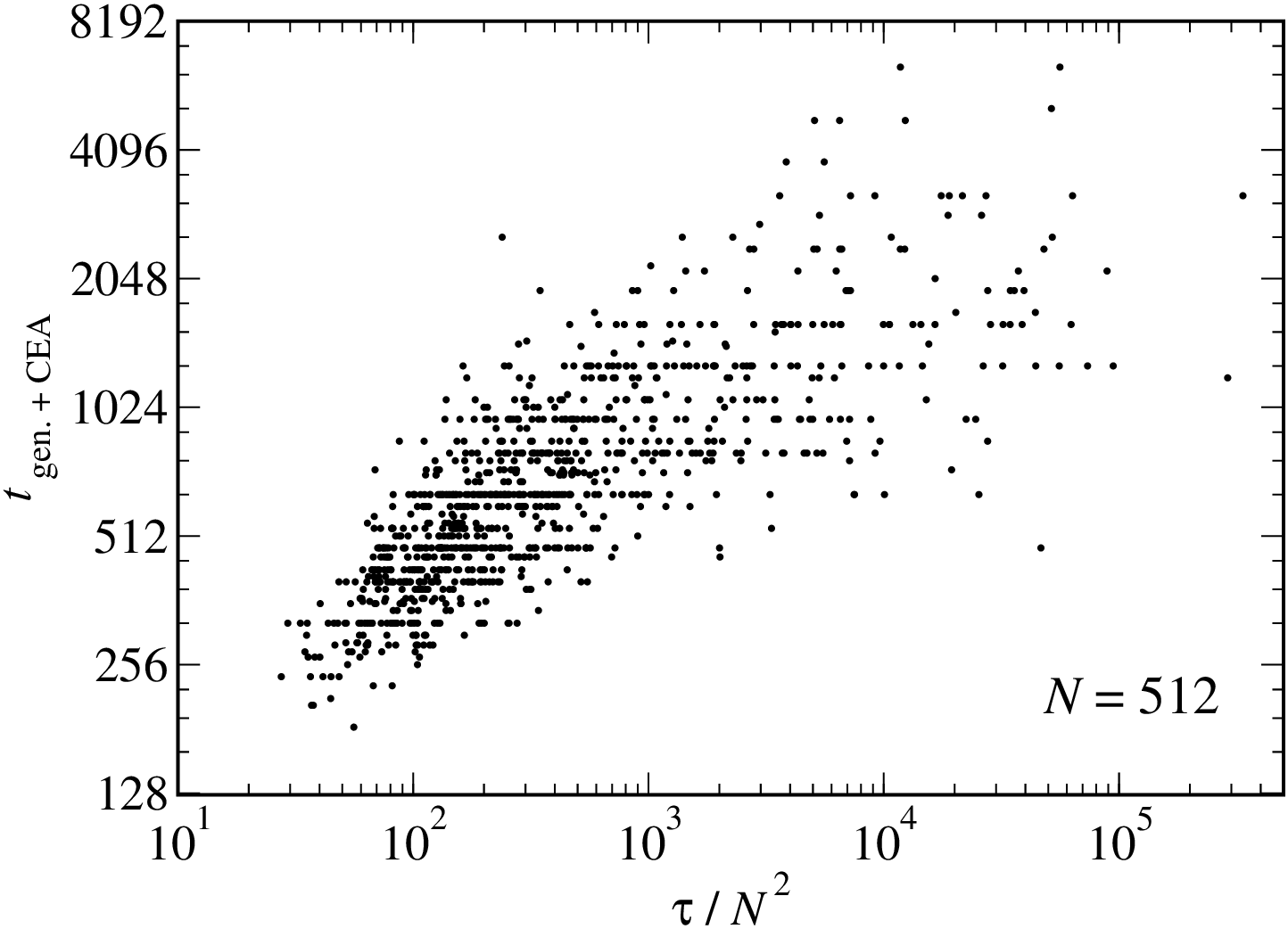}
  \caption{
    Correlation of the computational effort, $t_{\rm gen. + CEA}$, of the genetic CEA (see text) 
    and the round-trip time $\tau/N^2$ measured for multicanonical 
    sampling in the final step of the WL algorithm for a system with
    $N=8^3$ spins (1000 samples).}
  \label{fig:correlationCEA_WL}
\end{figure}

In Fig.~\ref{fig:correlationCEA} the correlation of the computational effort of the genetic CEA algorithm is shown versus the ratio $R_2$ of the density of states as defined above for 1000 samples with $N=8^3=512$ spins. 
There is only a weak correlation indicating that the heuristic algorithm is less sensitive to the energy-landscape close to the ground state than the WL algorithm.
A direct comparison between the genetic CEA and the WL algorithm is shown in 
Fig.~\ref{fig:correlationCEA_WL}. We find that the computational effort for the heuristic algorithm spreads over only 2 orders of magnitude in comparison to some 4 orders of magnitude for the WL algorithm. The correlation between the two algorithms is weak, and less pronounced for those samples which are especially hard to equilibrate using multicanonical sampling.

\subsection{Scaling of typical round-trip times}

\begin{figure}
  \includegraphics[scale=0.35]{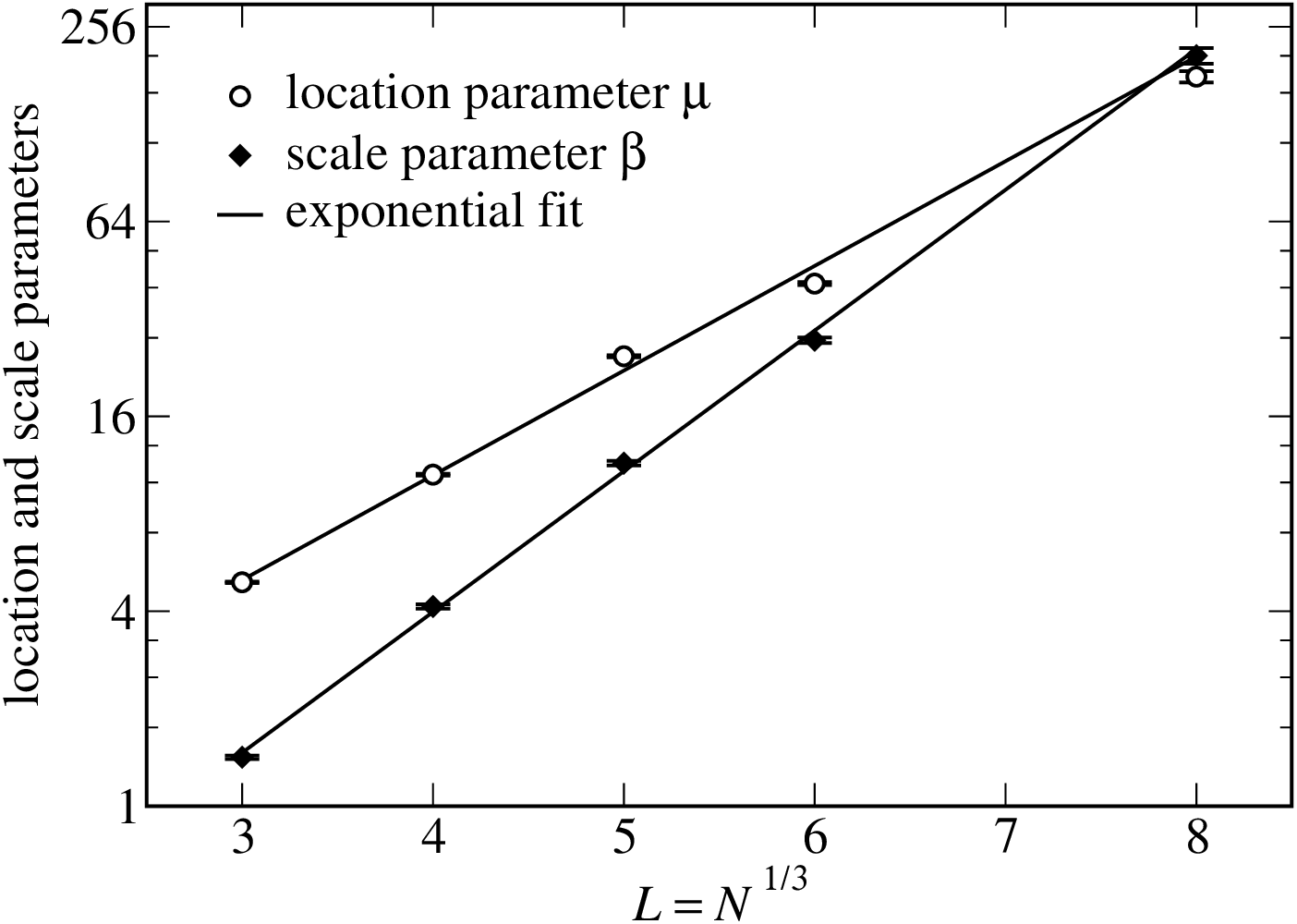}
  \caption {
    Scaling of the location ($\mu$) and scale ($\beta$) parameter of the Fr\'echet distribution
    versus the linear system size $L=N^{1/3}$. The parameters were fitted by a maximum-likelihood 
    estimator.
    The solid lines indicate exponential fits of the respective data points.  }
  \label{fig:beta_mu}
\end{figure}

Although the mean round-trip times are no longer well-defined for larger systems, the location and scale parameter of the Fr\'echet distribution stay well-defined and can be used to further characterize the scaling of these fat-tailed distributions with system size. 
Fig.~\ref{fig:beta_mu} shows that these parameters exponentially diverge with linear system size like
\begin{eqnarray}
  \mu &\propto& \exp \left( L/(1.34 \pm 0.03) \right) \;, \nonumber \\
  \beta &\propto& \exp \left( L/(1.00 \pm 0.03) \right) \;.
\end{eqnarray}
The study of the equilibrium behavior of larger system size is thus not only limited by the occurrence of rare events, but also by the exponential growth of the round-trip times in the ``bulk'' of the distribution which renders a comprehensive study of the sample-to-sample variations impossible.


\section{Diffusivity measurements}
\label{Diffusivity}

We have seen that both the asymptotic and the dynamic performance are
limited by the rough energy landscape close to the ground-state. To
study these limitations in more detail we measure the local
diffusivity of the random walker in energy. Recently, it was shown
that the simulated statistical ensemble can be optimized by a feedback
loop which reweights the ensemble based on preceeding measurements of
the local diffusivity.\cite{Huse:04} Although we do not follow up on this idea in the present study, we can measure the diffusivity to reveal the ``bottlenecks" of the biased random walk in energy as local minima in the diffusivity. Here we use a time-dependent definition of the diffusivity, $D(E,t_D)$, in energy space
\begin{equation}
  D(E,t_D) = \left< \left( E(t)-E(t+t_D) \right)^2 \right> / t_D \;,
  \label{eq:diffusivity}
\end{equation}
where $t_D$ is the diffusion time. The relevant time scale for the diffusion time is set by the round-trip time in energy.

\begin{figure}
  \includegraphics[scale=0.35]{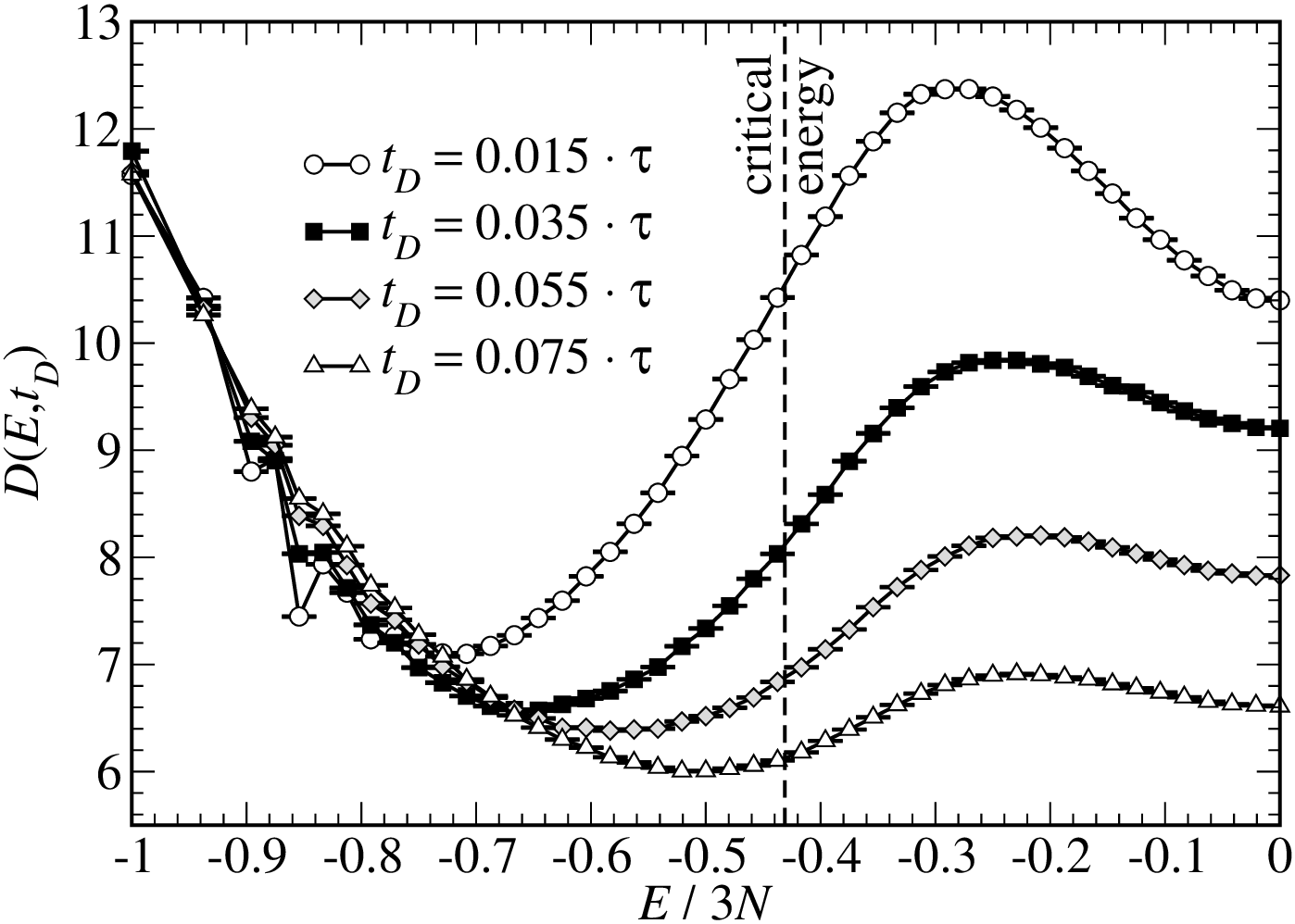}
  \caption{
    Local diffusivity $D(E,t_D)$ in energy of a flat-histogram random walker for the three-dimensional 
    Ising ferromagnet. Data for various diffusion times are shown for
    a system with $N=4^3=64$ spins.
  }
  \label{fig:diffusivity_fm}
\end{figure}

As a first example we study the local diffusivity of the
three-dimensional ferromagnetic Ising model which undergoes a
second-order phase transition from a magnetically ordered to a
disordered phase at finite energy $E_c / 3N \cong -0.43$.\cite{Ballesteros:98} The measured diffusivity $D(E,t_D)$ of the flat-histogram random walker is shown in Fig.~\ref{fig:diffusivity_fm}.
We find that the diffusivity is not constant as it is expected for an unbiased random walk, but there is a broad minimum below the critical energy. In this energy region the random walker is slowed down due to the slow dynamics of domain walls which separate droplets of magnetically ordered phases.

\begin{figure}
  \includegraphics[scale=0.35]{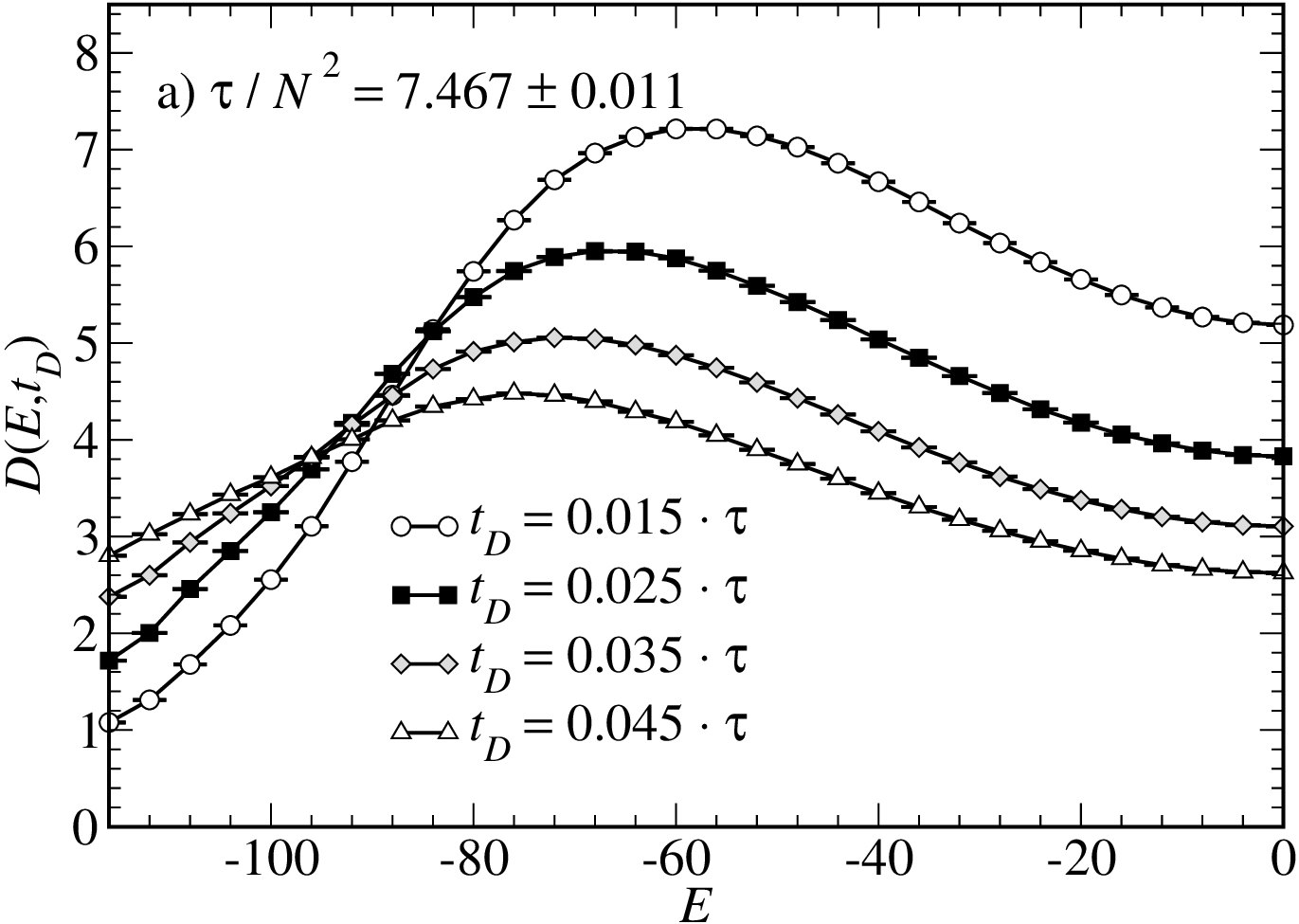}
  \includegraphics[scale=0.35]{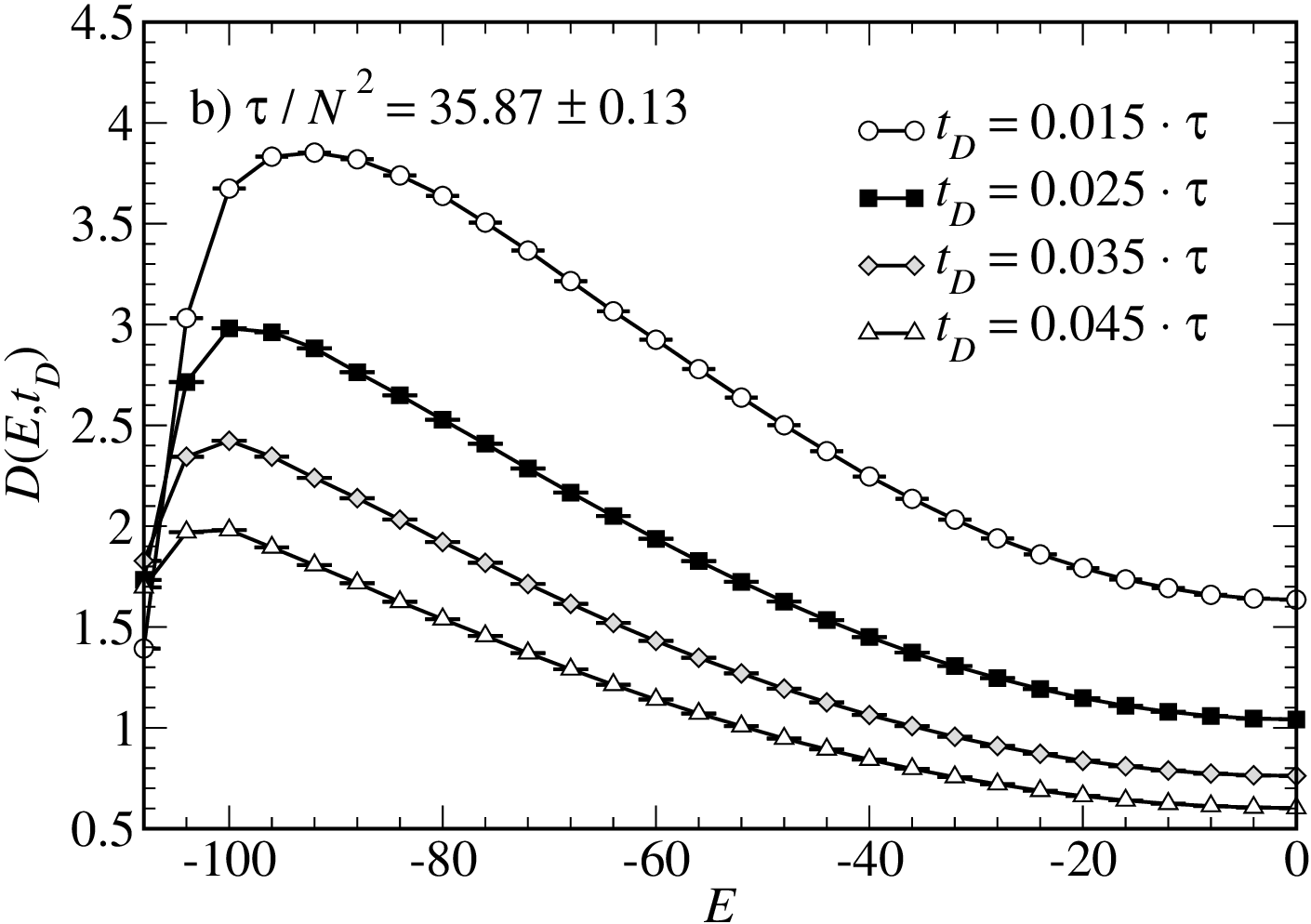}
  \caption{
    Local diffusivity $D(E,t_D)$ in energy of a flat-histogram random walker for two three-dimensional 
    $\pm J$ Ising spin glass samples. 
    Data for various diffusion times are shown for systems with
    $N=4^3=64$ spins.  }
  \label{fig:diffusivity_sg}
\end{figure}

Next we turn to the three-dimensional spin glass. Measurements of the diffusivity of the flat-histogram random walker for two randomly generated samples are shown in Fig.~\ref{fig:diffusivity_sg}. For both samples we find a minimum of the diffusivity at the ground-state energy.  We further note that with increasing round-trip times the minimum in the diffusivity becomes more pronounced. 
These diffusivity measurements further underline that the bottleneck of the flat-histogram random walker are at the ground-state energy. The suppressed diffusivity in this energy region gives rise to an entropic barrier which results in long round-trip times and aggravates equilibration of the random walker in the low energy phase. For larger systems the suppressed local diffusivity of the flat-histogram random walker renders a comprehensive study of the glassy phase impossible as it becomes computationally too expensive to equilibrate a large number of samples needed to calculate statistical averages.
The recent development of a systematic means to optimize the simulated statistical ensemble\cite{Huse:04} holds promise to overcome some of these limitations.

\section{Conclusions}

To summarize, we have studied the performance of the Wang-Landau
algorithm for the three-dimensional $\pm J$ Ising spin glass. The
asymptotic performance --- which corresponds to the performance of any
flat-histogram method sampling a multicanonical ensemble such as the
multicanonical method,\cite{Multicanonical} simulated and parallel
tempering,\cite{Tempering} broad histograms\cite{BroadHistograms} and
transition matrix Monte Carlo\cite{TMMC} --- is found to be dominated by strong sample-to-sample variations. 
The measured round-trip times follow fat-tailed Fr\'echet extremal-value distributions. The typical round-trip times in the bulk of these distributions obey exponential scaling, as expected for this NP-hard problem. The fat tails of the distributions dominate the calculation of statistical averages. A careful statistical analysis is needed which goes beyond the calculation of the moments of a finite sample distribution as it was done in previous studies.
The intrinsic character of the Fr\'echet extremal distributions becomes evident in strong correlations over several orders of magnitude between the round-trip time and the behavior of the density of states near ground-state energy. The origin of the extremal character of every single spin glass sample remains an open question which deservers further investigations.

Our measurements of the  dynamic performance and comparison with
ground states obtained by genetic CEA showed that for
 samples with up to $N=8^3$ spins the Wang-Landau
algorithm always finds the correct ground-state energies. 
For sizes $10^3$, $12^3$, one can find true ground
states, if one restricts the random walk to a small energy bin around the 
exact ground-state energy calculated by the heuristic CEA approach.
For samples with $N=14^3$ spins we identified samples for which the
Wang-Landau algorithm does not find the true ground-state energy
within reasonable simulation times ($\approx 10^7$ MC sweeps) and does
not converge towards the multicanonical ensemble.

The entropic barrier near the zero-temperature ground state can be
revealed as a pronounced minimum in the local diffusivity of the
flat-histogram random walker in energy. The recent development of
optimized statistical ensembles based on feedback of the local
diffusivity holds promise to overcome the observed slow down of the
flat-histogram random walker and thereby enhancing equilibration 
in these systems. \cite{Huse:04} 


\section{Acknowledgments}
We thank F.~Alet, H.~Gould, D.~A.~Huse, J.~Machta, and S.~Wessel for discussions. 
Part of the numerical calculations were performed on the Asgard Beowulf cluster at ETH
Z\"urich. ST and MT were supported by the Swiss National Science Foundation. ST and MT acknowledge hospitality of the Kavli Institute for Theoretical Physics at UC Santa Barbara which is supported by U. S. National Science Foundation Grant No. PHY99-07949. MT further acknowledges support of the Aspen Center for Physics. AKH was supported by the {\em VolkswagenStiftung} (Germany) within the program ``Nachwuchsgruppen an Universit\"aten''.

\bibliography{paper}

\begin{thebibliography}{12}
\expandafter\ifx\csname natexlab\endcsname\relax\def\natexlab#1{#1}\fi
\expandafter\ifx\csname bibnamefont\endcsname\relax
  \def\bibnamefont#1{#1}\fi
\expandafter\ifx\csname bibfnamefont\endcsname\relax
  \def\bibfnamefont#1{#1}\fi
\expandafter\ifx\csname citenamefont\endcsname\relax
  \def\citenamefont#1{#1}\fi
\expandafter\ifx\csname url\endcsname\relax
  \def\url#1{\texttt{#1}}\fi
\expandafter\ifx\csname urlprefix\endcsname\relax\def\urlprefix{URL }\fi
\providecommand{\bibinfo}[2]{#2}
\providecommand{\eprint}[2][]{\url{#2}}

\bibitem{SpinGlassReview}
    K. Binder and A.P. Young, Rev. Mod. Phys. {\bf 58}, 801 (1986);
\bibinfo{author}{\bibnamefont{M.~M{\'e}zard}, \bibnamefont{G.~Parisi},
  \bibnamefont{and} \bibnamefont{M.~A.~Virasoro}},
  \emph{\bibinfo{title}{Spin glass theory and beyond}},
  \bibinfo{publisher}{World Scientific, Singapore}
  (\bibinfo{year}{1987}); 
K.H. Fisher and J.A. Hertz, {\em Spin Glasses}\/, (Cambridge
University Press, Cambridge 1991); 
A.P. Young (ed.), {\em Spin glasses and random fields}\/, (World Scientific,
    Singapore 1998).

\bibitem{Ballesteros:00}
\bibinfo{author}{\bibnamefont{H.~G.~Ballesteros}, \bibnamefont{A.~Cruz}, \bibnamefont{L.~A.~Fern{\'a}ndez}, \bibnamefont{V.~Mart{\'i}n-Mayor}, \bibnamefont{J.~Pech}, \bibnamefont{J.~J.~Ruiz-Lorenzo}, \bibnamefont{A.~Taranc{\'o}n}, \bibnamefont{P.~T{\'e}llez}, \bibnamefont{C.~L.~Ullod}, \bibnamefont{and} \bibnamefont{C.~Ungil}},
  \bibinfo{journal}{Phys. Rev. B} \textbf{\bibinfo{volume}{62}},
  \bibinfo{pages}{14237} (\bibinfo{year}{2000}).

\bibitem{Palassine:99}
\bibinfo{author}{\bibnamefont{M.~Palassini} \bibnamefont{and} \bibnamefont{S.~Caracciolo}},
  \bibinfo{journal}{Phys. Rev. Lett.} \textbf{\bibinfo{volume}{82}},
  \bibinfo{pages}{5128} (\bibinfo{year}{1999}).

\bibitem{Multicanonical}
\bibinfo{author}{\bibfnamefont{B.~A.} \bibnamefont{Berg}} \bibnamefont{and}
  \bibinfo{author}{\bibfnamefont{T.}~\bibnamefont{Neuhaus}},
  \bibinfo{journal}{Phys. Rev. Lett.} \textbf{\bibinfo{volume}{{\bf 68}}},
  \bibinfo{pages}{9} (\bibinfo{year}{1992});
  \bibinfo{journal}{Phys. Lett. B.} \textbf{\bibinfo{volume}{{\bf 267}}},
  \bibinfo{pages}{249} (\bibinfo{year}{1991}).

\bibitem{Tempering}
\bibinfo{author}{\bibnamefont{K.~Hukushima} \bibnamefont{and} \bibfnamefont{K.~Nemoto}},
  \bibinfo{journal}{J. Phys. Soc. Japan} \textbf{\bibinfo{volume}{{\bf 65}}},
  \bibinfo{pages}{1604} (\bibinfo{year}{1996});
\bibinfo{author}{\bibfnamefont{E.~Marinari} \bibfnamefont{and} \bibnamefont{G.~Parisi}},
  \bibinfo{journal}{Europhys. Lett.} \textbf{\bibinfo{volume}{{\bf 19}}},
  \bibinfo{pages}{451} (\bibinfo{year}{1992});
\bibinfo{author}{\bibnamefont{A.~P.~Lyubartsev}, \bibnamefont{A.~A.~Martsinovski}, \bibnamefont{S.~V.~Shevkunov}, \bibnamefont{and} \bibnamefont{P.~N.~Vorontsov-Velyaminov}},
  \bibinfo{journal}{J.~Chem.~Phys.}
  \textbf{\bibinfo{volume}{{\bf 96}}}, \bibinfo{pages}{1776}
  (\bibinfo{year}{1992}).

\bibitem{BroadHistograms}
\bibinfo{author}{\bibfnamefont{P.~M.~C.~de~Oliveira}, \bibnamefont{T.~J.~P.~Penna}, \bibnamefont{and} \bibnamefont{H.~J.~Herrmann}},
  \bibinfo{journal}{Braz. J. Phys.} \textbf{\bibinfo{volume}{{\bf 26}}},
  \bibinfo{pages}{677} (\bibinfo{year}{1996}).

\bibitem{TMMC}  
\bibinfo{author}{\bibfnamefont{J.~S.~Wang}, \bibnamefont{T.~K.~Tay}, \bibnamefont{and} \bibnamefont{R.~H.~Swendsen}},
  \bibinfo{journal}{Phys. Rev. Lett.} \textbf{\bibinfo{volume}{{\bf 82}}},
  \bibinfo{pages}{476} (\bibinfo{year}{1999});
\bibinfo{author}{\bibfnamefont{J.~S.~Wang and R.~H.~Swendsen}},
  \bibinfo{journal}{J. Stat. Phys.} \textbf{\bibinfo{volume}{{\bf 106}}},
  \bibinfo{pages}{245} (\bibinfo{year}{2001}).

\bibitem{WangLandau}
\bibinfo{author}{\bibfnamefont{F.}~\bibnamefont{Wang}} \bibnamefont{and}
  \bibinfo{author}{\bibfnamefont{D.~P.} \bibnamefont{Landau}},
  \bibinfo{journal}{Phys. Rev. Lett.} \textbf{\bibinfo{volume}{{\bf 86}}},
  \bibinfo{pages}{2050} (\bibinfo{year}{2001}{\natexlab{a}});
  \bibinfo{journal}{Phys. Rev. E} \textbf{\bibinfo{volume}{{\bf 64}}},
  \bibinfo{pages}{056101} (\bibinfo{year}{2001}{\natexlab{b}}).

\bibitem{Barahona:82}
\bibinfo{author}{\bibfnamefont{F.}~\bibnamefont{Barahona}},
  \bibinfo{journal}{J.\ Phys.\ A} \textbf{\bibinfo{volume}{15}},
  \bibinfo{pages}{3241} (\bibinfo{year}{1982}).

\bibitem{Bieche:80}
\bibinfo{author}{\bibfnamefont{I.~Bieche}, \bibnamefont{R.~Maynard}, \bibnamefont{R.~Rammal}, \bibnamefont{and} \bibnamefont{J.~P.~Uhry}},
  \bibinfo{journal}{J. Phys. A} \textbf{\bibinfo{volume}{13}},
  \bibinfo{pages}{2553} (\bibinfo{year}{1980}).

\bibitem{SaulKardar:94}
\bibinfo{author}{\bibfnamefont{L.~K.} \bibnamefont{Saul}} \bibnamefont{and}
  \bibinfo{author}{\bibfnamefont{M.}~\bibnamefont{Kardar}},
  \bibinfo{journal}{Nuclear Physics B} \textbf{\bibinfo{volume}{432}},
  \bibinfo{pages}{641} (\bibinfo{year}{1994}).

\bibitem{simone95} C. De Simone, M. Diehl, M. J\"unger, P. Mutzel, 
G. Reinelt, and G. Rinaldi, {\it J. Stat. Phys.}\/ {\bf 80}, 487 (1995).

\bibitem{simone96} C. De Simone, M. Diehl, M. J\"unger, P. Mutzel, 
G. Reinelt, and G. Rinaldi, {\it J. Stat. Phys.}\/ {\bf 84}, 1363 (1996).

\bibitem{Hartmann}
\bibinfo{author}{\bibfnamefont{A.~K.~Hartmann}},
  \bibinfo{journal}{Physica A} \textbf{\bibinfo{volume}{224}},
  \bibinfo{pages}{480} (\bibinfo{year}{1996});
  \bibinfo{journal}{Phys. Rev. E} \textbf{\bibinfo{volume}{59}},
  \bibinfo{pages}{84} (\bibinfo{year}{1999}).

\bibitem{Hartmann:Book}
\bibinfo{author}{\bibnamefont{A.~K.~Hartmann} \bibnamefont{and} \bibnamefont{H.~Rieger}},
  \emph{\bibinfo{title}{Optimization Algorithms in Physics}},
  \bibinfo{publisher}{Wiley-VCH}, \bibinfo{address}{Berlin}
  (\bibinfo{year}{2001}).

\bibitem{Dayal:04}
\bibinfo{author}{\bibfnamefont{P.~Dayal}, \bibnamefont{S.~Trebst}, \bibnamefont{S.~Wessel}, \bibnamefont{D.~W{\"u}rtz}, \bibnamefont{M.~Troyer}, \bibnamefont{S.~Sabhapandit}, \bibnamefont{and} \bibnamefont{S.~N.~Coppersmith}},
  \bibinfo{journal}{Phys. Rev. Lett. {\bf 92}, 097201}
  (\bibinfo{year}{2004}).

\bibitem{Extremal1}
\bibinfo{author}{\bibfnamefont{P.}~\bibnamefont{Embrechts}},
  \bibinfo{author}{\bibfnamefont{C.}~\bibnamefont{Kl\"upelberg}},
  \bibnamefont{and} \bibinfo{author}{\bibfnamefont{T.}~\bibnamefont{Mikosch}},
  \emph{\bibinfo{title}{Modelling Extremal Events}},
  \bibinfo{publisher}{Springer, Berlin} (\bibinfo{year}{1997}).

\bibitem{Extremal2}
\bibinfo{author}{\bibfnamefont{R.~A.} \bibnamefont{Fisher}} \bibnamefont{and}
  \bibinfo{author}{\bibfnamefont{L.~H.~C.} \bibnamefont{Tippett}},
  \bibinfo{journal}{Proc. Cambridge Philos. Soc.}
  \textbf{\bibinfo{volume}{{\bf 24}}}, \bibinfo{pages}{180}
  (\bibinfo{year}{1928}).

\bibitem{Extremal3}
\bibinfo{author}{\bibnamefont{R.-D.~Reiss} \bibnamefont{and}
\bibnamefont{M.~Thomas}},
  \emph{\bibinfo{title}{Statistical Analysis of Extreme Values}},
  \bibinfo{publisher}{Birkh{\"a}user, Basel} (\bibinfo{year}{2001}).

\bibitem{footnote}
\bibnamefont{For the {\em finite} set of $2^{2N}$ distinct spin glass realizations of a given system size $N$ the Fr\'echet distribution is cutoff at some finite value. The average round-trip time does not diverge, but is dominated by this cutoff value. For the system with $N = 8^3$ spins we can estimate the average round-trip time form the fitted Fr\'echet distributions to be $\tau_{\text{average}} > 10^{100}$ which for all practical purposes cannot be distinguished from a divergence.}

\bibitem{Schulz:03}
\bibinfo{author}{\bibfnamefont{B.~J.~Schulz}, \bibnamefont{K.~Binder}, \bibnamefont{M.~M{\"u}ller}, and \bibnamefont{D.~P.~Landau}},
  \bibinfo{journal}{Phys. Rev. E} \textbf{\bibinfo{volume}{67}},
  \bibinfo{pages}{067102} (\bibinfo{year}{2003}).

\bibitem{Huse:04}
\bibinfo{author}{\bibnamefont{D.~A.~Huse}, \bibnamefont{S.~Trebst}, \bibnamefont{and} \bibnamefont{M.~Troyer}},
  \bibnamefont{cond-mat/0401195}.

\bibitem{Ballesteros:98}
\bibinfo{author}{\bibnamefont{H.~G.~Ballesteros}, \bibnamefont{L.~A.~Fern{\'a}ndez}, \bibnamefont{V.~Mart{\'i}n-Mayor}, \bibnamefont{A.~Mu{\~n}oz~Sudupe}, \bibnamefont{G.~Parisi}, \bibnamefont{and} \bibnamefont{J.~J.~Ruiz-Lorenzo}},
  \bibinfo{journal}{J. Phys. A} \textbf{\bibinfo{volume}{32}},
  \bibinfo{pages}{1} (\bibinfo{year}{1999}).

\end{thebibliography}

\end{document}